\documentclass[aps,pra,twocolumn,superscriptaddress]{revtex4}

\usepackage{color,epsfig,graphicx}
\usepackage{amssymb,bbm}
\usepackage{amsmath}

\def\Id{\mathbbm{1}}
\def\rme{\mathrm{e}}
\def\ket#1{\left|#1\right>}
\newcommand{\Tr}{\mathop{\mathrm{Tr}}\nolimits}
\def\CNOT{\textsc{cnot}}
\def\SWAP{\textsc{swap}}
\def\NOT{\textsc{not}}

\begin{document}

\title{Vacuum induced Stark shifts for quantum logic using a
  collective system in a high quality dispersive cavity}

\author{A. G\'abris}
\affiliation{
Research Institute of Solid State Physics and Optics,
Hungarian Academy of Sciences\\
H-1525 Budapest P.O. Box 49, Hungary}
\affiliation{
Physical Research Laboratory,
Navrangpura, Ahmedabad -- 380009, India}
\author{G.~S. Agarwal}
\affiliation{
Department of Physics, Oklahoma State University\\
Stillwater, OK 74078, USA}
\affiliation{
Physical Research Laboratory,
Navrangpura, Ahmedabad -- 380009, India}

\date{\today}


\begin{abstract}
  A collective system of atoms in a high quality cavity can be
  described by a nonlinear interaction which arises due to the Lamb
  shift of the energy levels due to the cavity vacuum [Agarwal et al.,
  Phys. Rev. A \textbf{56}, 2249 (1997)]. We show how this
  collective interaction can be used to perform quantum logic. In
  particular we produce schemes to realize \CNOT{} gates not only for
  two-qubit but also for three-qubit systems. We also discuss
  realizations of Toffoli gates. Our effective Hamiltonian is also
  realized in other systems such as trapped ions or magnetic
  molecules.
\end{abstract}

\maketitle

\section{Introduction}

The possibility of doing quantum computation with neutral atoms is
becoming more realistic with the advances in techniques relating to
the trapping of few atoms which could even be addressed
individually\cite{Haroche-prl93,Reichel-nature413,Kimble-prl90}.
However a number of experiments so far have been done with flying
qubits\cite{Rauschenbeutel-prl83, Kimble-prl75, Haroche-pra52} and a
number of proposals exist on implementing quantum logic operations
using cavity QED\cite{Pellizzari-prl75, Haroche-jmo50, Agarwal-pra69,
  Scully-pra65, Agarwal-pra70}. We note that the realization of a
\CNOT{} gate between two qubits requires some form of interaction
between the qubits.  There are thus realizations which depend on the
interaction between the center of mass degrees and the electronic
degrees of freedom as in the case of ions\cite{Wineland-nature429,
  Blatt-nature429, Wineland-prl89}, the interaction between the
photonic qubit and the atom as in case of cavity QED
\cite{Rauschenbeutel-prl83}. Thus for doing logic operations with
neutral atoms one would require an effective interaction between them.
Note that we have to keep the distance between atoms such that
selective addressing is possible for one qubit operations. On the
other hand if atoms are far apart then the electrostatic interaction
between them is very weak. These problems can be overcome by using a
high quality dispersive cavity.  It has been shown earlier that the
interaction of trapped atoms with a single mode of the radiation field
produces an effective interaction which can be utilized for doing
quantum logic\cite{Agarwal-pra56}. Though we shall work in the
framework of this physical system, it is notable that the considered
Hamiltonian is a special case of the Lipkin model\cite{Lipkin-np62},
and similar Hamiltonians can be associated with the dynamics of ion
traps\cite{ Fleischhauer-prl90} and Fe$^{3+}$ ions of a large magnetic
molecule\cite{Garg-prb63}.

The outline of this paper is as follows. We shall introduce our system
and qubits in section \ref{sec:system}, present a brief summary of
some key mathematical tools used during our calculations in section
\ref{sec:eng2qubit}, then in section \ref{sec:N2} and \ref{sec:N3} we
shall give specific constructions of \CNOT{} gates for $N=2$ and $N=3$
atoms, respectively. We discuss realizations of Toffoli gates in
section \ref{sec:toffoli}, and section \ref{sec:conclusions} is
dedicated to our conclusions. We note that we work strictly with
qubits and do not use any additional levels for the logic
operation\cite{Cirac-prl79}. The use of additional
levels\cite{LYou-prl90, Solano-pra64, Feng-pra66, Jane-pra65,
  Walther-prl89, Yang-pra70, LYou-pra67} may lead to additional
sources of decoherence either due to local environment or due to the
fields which are used to address such levels.

\section{Physical system}
\label{sec:system}

We consider $N$ two-level atoms trapped in a cavity, with the atomic
transition frequency $\omega_0$ detuned from the cavity resonance
frequency $\omega$ by some value $\Delta$, and denote the dipole
coupling between an atom and the cavity by $g$. The cavity losses are
characterized by a constant $\kappa$ that describes the coupling to an
external reservoir. We do not consider atomic spontaneous decay
explicitly, but we note that for large enough detuning $\Delta$, the
modification of the decay rate due to the Purcell effect becomes
negligible. Further, in order to facilitate individual addressing of
the atoms, we require that the atoms are well separated, i.e.\ their
spatial wave functions are non-overlapping.

It was shown in Ref.~\cite{Agarwal-pra56} that if the cavity is in a
thermal state with mean photon number $\bar{n}$, tracing out for this
cavity mode in the limit $g\sqrt{N} \ll |i\Delta + \kappa|$ results in
a time evolution of the atoms that can well be approximated by a
unitary process. The effective Hamiltonian corresponding to this
evolution may be written most conveniently in terms of the $S_i$
collective spin-$N/2$ and $S^2$ total angular momentum square operators
defined through
\begin{equation}
S_i = \frac12 \sum_{k=1}^N \sigma^{(k)}_i,
\end{equation}
where the $\sigma^{(k)}_i$ operators are the Pauli-$i$ operators
($i=x,y,z$ or $+,-$) defined on the computational basis as usual. We
define the computational basis states $\ket0_k$ and $\ket1_k$ as the
ground ($\ket{g}$) and excited ($\ket{e}$) states of the $k$th atom,
respectively. The number of qubits is therefore equal to the number of
atoms $N$ trapped within the cavity. The effective Hamiltonian for $N$
atoms reads
\begin{subequations}
\label{eq:HS}
\begin{eqnarray}
H_N &=& \hbar \eta (S_+S_- + 2\bar{n} S_z) \label{eq:HSpSm} \\
  &=& \hbar \eta \left[ S^2 - S_z^2 + (2\bar{n}+1) S_z\right],
\label{eq:HS2}
\end{eqnarray}
\end{subequations}
where the coupling factor is $\eta=g^2\Delta/(\kappa^2 + \Delta^2)$.

The main theme of this paper shall be to prove the universality of
this interaction Hamiltonian. For simplicity we assume that 1-qubit
operations can be carried out much faster than the characteristic time
of the collective interaction. To prove universality we show that this
way it is possible to generate all \CNOT{} gates.

We note that the terms linear in $S_z$ in Hamiltonians (\ref{eq:HS})
correspond to 1-qubit operations. Let
\begin{equation}
R_{x,y,z}(\vartheta) =
\exp(-i\vartheta\sigma_{x,y,z}/2)
\end{equation}
denote to standard SU(2) rotations.  Since $S_z$ commutes with the
rest of the Hamiltonian, it follows that the linear terms in
Hamiltonians (\ref{eq:HSpSm}) and (\ref{eq:HS2}) may be effectively
cancelled from the time-evolution by applying $R_z(-2\eta\bar{n}t)$
and $R_z \left(-\eta(2\bar{n}+1)t\right)$, respectively, to every
qubit. It is important that the angle of rotation depends on the
actual mean photon number (i.e.\ temperature) of the cavity. It also
depends on the time $t$ for which $H_N$ is to be applied, however, as
it shall be shown, in the course of quantum logic gate operation, this
$t$ is known \`a priori. Also, because of the mentioned commutation
properties this rotation can be carried out any time within the
time-window prescribed by $t$. Later in this paper we shall work with
Hamiltonians (\ref{eq:HS}) without the linear terms, and assume that
the linear terms are always compensated with these 1-qubit rotations.

\section{Engineering two-qubit gates}
\label{sec:eng2qubit}

For construction of desired two-qubit gates we have used a technique
introduced in Ref.~\cite{Makhlin-qi1}. For conciseness we briefly
summarize this technique.

We consider two two-qubit gates $M$ and $L$, with unit determinants
for now. We term them equivalent if they can be transformed into each
other using only one-qubit operations $O=O_1 \otimes O_2$ and $O'=O'_1
\otimes O'_2$ as
\begin{equation}
L = O' M O.
\label{eq:equivalence}
\end{equation}
Here we used the tensorial product notation $\otimes$ to distinguish
operators acting on different subsystems. A very important result of
\cite{Makhlin-qi1} is that this equivalence is perfectly characterized
by two numbers: Let $M_B=Q^{\dag}M Q$ ($L_B$ similarly) with $Q$ being
the unitary transformation to a specific entangled basis related to
the standard Bell-states, and $m=M_B^T M_B$ ($l$ similarly) where $^T$
denotes real transpose. Then the pairs ($\Tr^2m$, $\Tr m^2$) and
($\Tr^2l$, $\Tr l^2$) coincide if and only if $L$ and $M$ are
equivalent in the sense of Eq.~(\ref{eq:equivalence}).

An interesting side-effect is that these matrices $m$ and $l$ can be
used to find the one-qubit operators $O$ and $O'$ connecting two
equivalent $M$ and $L$. The recipe is as follows: first find the
diagonal basis of $m$ ($l$), i.e.~ write $m=O_M^T d_M O_M$ where $d_M$
is a diagonal matrix. Then the solution can be written as
\begin{subequations}
\label{eq:1qb-eq}
\begin{eqnarray}
 O &=& O_M O_L^T \\
 O' &=& O_L^T O^T M_B^{\dag}.
\end{eqnarray}
\end{subequations}

These results can also be applied to matrices with non-unit
determinants, and to cover that case also we associate the numbers
\begin{equation}
[\Tr^2m/ 16\det M, (\Tr^2m-\Tr m^2)/ 4\det M]
\label{eq:invariants}
\end{equation}
with a matrix $M$. This pair of numbers can be viewed as invariants
under one-qubit operations.

The equations (\ref{eq:1qb-eq}) can be used to construct $L$ using $M$
only if $L$ and $M$ are equivalent. However, the invariants
(\ref{eq:invariants}) can also be used to construct $L$ from a given
matrix $A$ not belonging to the same equivalence class. The method
relies on splitting the problem into two, by first searching for a
matrix equivalent to our target $L$ in the form of $M=A O_f A$, and
then using (\ref{eq:1qb-eq}) to obtain $L$. Solving the invariant
equations is generally less involved than the solution of matrix
equations.  Unfortunately the first step on the course is not
guaranteed to work for all $L$ and $A$. For example, a \CNOT{} gate
(or any equivalent) may never be constructed from \SWAP{} gates and
1-qubit operations.  For some other $A$, the construction may be
possible, but only by using $A$ more than twice, e.g.\ $A O_f A O_f'
A$.

\section{Case of $N=2$ atoms}
\label{sec:N2}

\begin{figure*}
\includegraphics[scale=0.8]{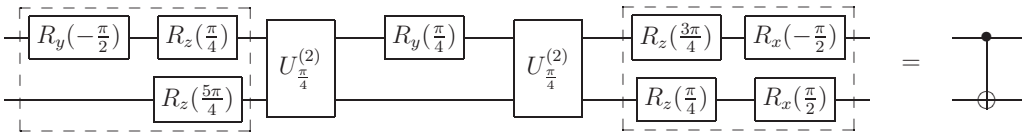}
\caption{Quantum circuit diagram depicting the sequence to prepare
  \CNOT{} gate from the time-evolution operator in
  (\ref{eq:U2}). 
}
\label{fig:CNOT2}
\end{figure*}

As the simplest case we consider two atoms in the cavity. The
collective spin operators now describe a spin-1 system, and a number
of simplifications apply to this case. For example, the Dicke states
\cite{Dicke-pr93,Arecchi-pra6} span the complete Hilbert space of the
two atoms and $G=[S^2 - (S_z^2 - S_z)]/2$ is a projector operator.
Considering (\ref{eq:HSpSm}) without the linear terms we have
$H_2=2\hbar \eta G$, and the time-evolution operator is
\begin{equation}
U^{(2)}(t) = e^{-\frac{i}{\hbar}H_2t}=1 - e^{-i\eta t} 2i G \sin(\eta t).
\end{equation}
In the computational basis this corresponds to the matrix
\begin{equation}
U^{(2)}(t) = e^{( - i\varphi )}\, \left( 
\begin{array}{cccc}
e^{ - i\varphi } & 0 & 0 & 0 \\
0 & \cos\varphi  &  - i\sin\varphi & 0 \\
0 &  - i\sin\varphi & \cos\varphi & 0 \\
0 & 0 & 0 & e^{i \varphi}
\end{array}
 \right),
\label{eq:U2}
\end{equation}   
with $\varphi=\eta t$. The invariants (\ref{eq:invariants}) of this
matrix are
\begin{equation}
[\cos^4 \varphi, 4\cos^2\varphi -1],
\end{equation}
while for a \CNOT{} gate we would require $[0,1]$. We see that this
requirement is not met by any real $\varphi$.  After some algebra,
however, we obtain that with $U^{(2)}_{\frac{\pi}4}=
U^{(2)}\left(\frac\pi4\eta^{-1}\right)$, the sequence
\begin{equation}
\tilde{U}^{(2)} = U^{(2)}_{\frac{\pi}4} O_f   
    U^{(2)}_{\frac{\pi}4},
\end{equation}
is equivalent to a \CNOT{} gate if $O_f=R_y(\pi/4)\otimes\Id$. In
particular, using this \CNOT{} equivalent gate $\tilde{U}^{(2)}$ a
\CNOT{} with first bit as control and second as target bit can be
produced as
\begin{equation}
\CNOT{} = \rme^{i\pi/4} O_c' \tilde{U}^{(2)} O_c, 
\label{eq:cnot-2}
\end{equation}
the one-qubit operations of this formula being
\begin{eqnarray}
O_c' &=& \left[ R_x(-\pi/2) R_z(3\pi/4) \right] \otimes 
 \left[ R_x(\pi/2) R_z(\pi/4) \right] \\
O_c  &=&  \left[ R_z(\pi/4) R_y(-\pi/2) \right] \otimes R_z(5\pi/4).
\end{eqnarray}
The phase factor is in principle irrelevant and is written there for
didactical reasons only. The construction is depicted as a quantum
circuit diagram on Fig.~\ref{fig:CNOT2}. We note here, that assuming
two-qubit gates $U^{(2)}$ with equal $t$, this construction is optimal
in terms of operation time for the complete \CNOT{} gate.

\section{Case of $N=3$ atoms}
\label{sec:N3}

To build quantum logic for three atoms using the Hamiltonian in
(\ref{eq:HS}) we first show how to construct \CNOT{} gates between any
two atoms. The reason for constructing first a two-qubit gate from the
three-atom collective interaction, rather than directly constructing a
universal three-qubit gate (e.g.\ Toffoli or Fredkin gate) is the lack
of convenient characterization of higher qubit gates.  Up to date,
invariants such as (\ref{eq:invariants}) have been discovered only
for two-qubit gates.

In this section we shall consider $H_3$ of (\ref{eq:HS2}) without
the linear terms, and using this we first construct a two-qubit gate
that connects only two atoms and leaves the third atom unchanged,
i.e.\ 
\begin{equation}
U^{(3)}_{23}= \Id \otimes U_{23}.
\label{eq:only-one}
\end{equation}
Following the scheme similar to the spin-echo technique, we search for
operators fulfilling (\ref{eq:only-one}) in the form
\begin{equation}
U^{(3)}_{23}(t)=X_1  U^{(3)}(t) X_1  U^{(3)}(t'),
\label{eq:spin-echo}
\end{equation}
where $U^{(3)}(t)=\exp({-\frac{i}{\hbar} t H_3})$ is the
time-evolution generated by the chosen Hamiltonian, and $X_1=R_x(\pi)
\otimes \Id \otimes \Id$ which is essentially a \NOT{} gate. We pose
the condition (\ref{eq:only-one}) on (\ref{eq:spin-echo}) to find the
appropriate $t$ and $t'$.

The time evolution operator $U^{(3)}(t)$ is diagonal in the
Dicke-state basis. To develop further insight into the problem, we
apply the theory of angular momentum addition, and separate our
spin-$3/2$ system into a product of a spin-$1/2$ and a spin-$1$
subsystem. Transformation matrix to the product basis is given by the
relevant Clebsch-Gordan coefficients.

We require that (\ref{eq:spin-echo}) acts on the spin-1/2 subsystem as
the identity, and this condition translates to $t'=t$ and
$\sin(3/2\eta t)=0$. This gives three distinct solutions for $U_{23}$
for each $i=-1,0,1$ via $\eta t=2/3 \pi (3k+i)$ $(k\in\mathbb{Z})$.
Out of these three, $i=0$ corresponds to the identity operator and is
therefore irrelevant. The solutions for $i=-1$ and $i=1$ are
essentially the same (adjoint of one another) and they both have
invariants $[1/4, 3/2]$. In the following we work out the \CNOT{} gate
explicitly for $i=1$, because its implementation requires less time.
This operation is represented by
\begin{equation}
U_{23} = \left( 
\begin{array}{cccc}
e^{-i\pi/3} & 0 & 0 & 0 \\
0 &  e^{i\pi/3} & 0 & 0 \\
0 & 0 &  e^{i\pi/3} & 0 \\
0 & 0 & 0 & e^{-i\pi/3}
\end{array} \right)
\label{eq:U23}
\end{equation}
in the computational basis. We note that (\ref{eq:U23}) may be written
as $\exp(-i \pi/3\, \sigma_z \otimes \sigma_z)$ resembling the
Heisenberg spin-spin interaction that has found many applications in
Quantum Information Processing, most notably NMR Quantum
Computing\cite{Chuang-nature393,Cory-phd120,AnilKumar-pra65}. However,
in this case the interaction time is fixed by the conditions on $t$
and $t'$ of (\ref{eq:spin-echo}). Nevertheless, it will be seen that
it is possible to express \CNOT{} gates using this operator and
1-qubit gates.

Having a well-defined two-qubit gate in hand we can again follow the
recipe of Sec.~\ref{sec:eng2qubit}. After some algebra we find that
using the one-qubit operators acting on the subspace of qubits 2 and
3,
\begin{eqnarray}
O_f &=& \Id \otimes R_y(\varphi_f)\\
O_c &=& R_x(-\pi/2) \otimes 
  [R_z(\pi) R_x(\varphi_c)]  \\
O_c' &=& [R_z(-\pi/2) R_y(-\pi)] \nonumber \\
&& \otimes [R_z(\varphi_c') R_y(-\pi/2) R_z(\pi/2)],
\label{eq:1qubit-3}
\end{eqnarray}
with
\begin{eqnarray}
\tan(\varphi_f/2)&=& 1/\sqrt2 \\
\tan(\varphi_c/2)&=& \sqrt{2/3}-1 \\
\tan(\varphi_c'/2)&=& (1-\sqrt3)/\sqrt2,
\end{eqnarray}
the \CNOT{} gate can be constructed as
\begin{equation}
\CNOT{} = \rme^{-i\pi/4} O_c'  U_{23} O_f U_{23} O_c.
\label{eq:cnot-3}
\end{equation}

This \CNOT{} gate acts on qubit $2$ and $3$ as control and target
bits, respectively. However, due to the symmetry of $H_3$, a \CNOT{}
gate acting the other way around or connecting different qubits is
achievable simply by exchanging the role of qubits appropriately.

We note here that while this implementation of \CNOT{} gates is exact,
it is not necessarily optimal, and the \CNOT{} may be realized on this
system more efficiently.

\section{Toffoli gates}
\label{sec:toffoli}

\begin{figure}
\includegraphics[scale=0.8]{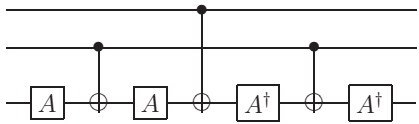}
\caption{Quantum circuit diagram for the simplified Toffoli
  gate requiring only three \CNOT{} gates\cite{Barenco-pra52}.
  ($A=R_y(\pi/4)$)}
\label{fig:toffoli}
\end{figure}

Universality of \CNOT{} gates implies that having them in all
configurations for three qubits allows the construction of any
three-qubit quantum gate, i.e.\ any SU($2^3$) operator.  As an example
we consider another important building block for systematic
construction of complex quantum circuits, the Toffoli
gate\cite{Barenco-pra52}. We also discuss a simplified version of the
Toffoli gate (Fig.~\ref{fig:toffoli}) that differs from the Toffoli
gate only in one conditional phase shift whereas requiring only half
the \CNOT{} gates. In some circumstances Toffoli gates may be replaced
by the simplified versions.

Using the expressions for \CNOT{} gates in our three-atom system, it
is straight-forward to implement both of these important quantum
gates. Simple arithmetic counting the number of applications of
$U^{(3)}(2\pi/3\eta^{-1})$ operations shows that the gate times for
the Toffoli and its simplified version add up to $16\pi/\eta$ and
$8\pi/\eta$, respectively. Following DiVincenzo's
criteria\cite{DiVincenzo-fortschr48}, for efficient error-free quantum
computation these gate times should be much shorter than the coherence
time of the complete system.

\section{Conclusions}
\label{sec:conclusions}

In this paper we have shown the computational universality upto three
qubits of a cavity assisted interaction between two-level atoms
trapped in a dissipative cavity. In addition to the collective
interaction we only needed single-qubit operations to implement
multi-qubit gates. This required that the atoms are separately
addressable, and we also assumed that single-atom operations can be
performed on much shorter time scales than the collective interaction.
The formalism used is not specific to two or three atom systems and
therefore allows for further generalizations to more qubits.

In comparison with earlier proposals, our scheme is not only robust to
cavity decay, but also allows dealing with thermal cavity states in a
straight-forward manner. We believe therefore that this scheme may
find useful applications in situations where good localization of
atoms had been achieved but the possibility of constructing good
cavities is limited.

\begin{acknowledgments}
  This work was partly supported by the National Research Fund of
  Hungary under contract Nos.\ T~034484 and T~043079; the Marie Curie
  Programme of the European Commission; and by the Hungarian Ministry
  of Education under contract No.\ CZ-5/03.
  A.~G.\ would also like to thank the Director of
  P.R.L., Ahmedabad for his warm welcome and for all his support, and
  J.\ Janszky for useful discussions.

\end{acknowledgments}


\end{document}